\documentclass[aps,prl,showpacs]{revtex4} 
\usepackage{amsmath} 
\usepackage{amssymb} 
\usepackage{graphicx} 
\usepackage{graphics}
\usepackage{epsfig}
\usepackage{slashed}
\newcommand\ba{\begin{eqnarray}}
\newcommand\ea{\end{eqnarray}}
\newcommand\alb{\begin{align}}
\newcommand\ale{\end{align}}
\newcommand\be{\begin{equation}}
\newcommand\ee{\end{equation}}

\usepackage{amsmath} 
\usepackage{amssymb}

\begin{document}

		\title{The fourth dimension of the nucleon structure: 
spacetime analysis of the timelike electromagnetic proton form factors} 


		\author{Andrea Bianconi} 
    \affiliation{\it Dipartimento di
	Ingegneria dell\!~$^\prime$Informazione, \\
Universit\`a degli Studi di Brescia, via Branze 38, I-25123 Brescia, Italy,\\
	and Istituto Nazionale di Fisica Nucleare, Sezione di Pavia,\\ 
via Bassi 6, Pavia, Italy}

		\author{Egle~Tomasi-Gustafsson} 
		\affiliation{\it IRFU, CEA, Universit\'e Paris-Saclay, 91191 Gif-sur-Yvette Cedex, France}

		\date{\today}

		\begin{abstract}
As well known, spacelike proton form factors expressed in 
the Breit frame may be 
interpreted as the Fourier transform of static space distributions of 
electric charge and current. In particular, the electric form 
factor is simply the Fourier transform of the charge distribution 
$F(q)=\int e^{i\vec q \cdot \vec r} \rho(r)d^3r$. 
We don't have an intuitive interpretation of 
the same level of simplicity for the proton timelike form factor  
appearing in the reactions $e^+e^-\leftrightarrow \bar{p}p$. However, 
one may suggest that in the center of mass (CM) frame, where 
$q_\mu x^\mu =qt$, a timelike electric form factor 
is the Fourier transform 
$F(q) =\int e^{iqt} R(t)dt$ of a function $R(t)$ 
expressing how the electric properties of the forming (or annihilating) 
proton-antiproton pair evolve in time. Here we analyze in depth this idea, show 
that the functions $\rho(r)$ and $R(t)$ can be formally
written as the time and space 
integrals of a unique correlation function depending on both time and space coordinates. 
                \end{abstract}
\maketitle

\section{Introduction}

\subsection{Background} 
The reaction $e^+ + e^-\to \bar p +p$ and its time reverse 
$\bar p + p \to e^+ + e^-$ 
have been used to extract the electromagnetic form factors (FFs)
of the proton in the time-like (TL) region. Assuming that the 
interaction occurs through one photon exchange, the annihilation cross 
section is expressed in terms of the FF moduli 
squared (\cite{Zichichi:1962ni},  see also \cite{Pacetti:2015iqa,Denig:2012by} for 
recent reviews on TLFF). 

The empirical 
knowledge and the theoretical understanding of the TLFF 
are less advanced than 
for the spacelike (SL) case.  In 
particular, an experimental separation of the electric and 
the magnetic FF has not been possible in the TL region, because of 
the available limited luminosity. The cross section $\sigma$ of the above reactions  
allows to extract the squared modulus of a single effective 
form factor $F_p$ \cite{Bardin:1994am}
\be
|F_p|^2=\displaystyle\frac{3\beta q^2 \sigma}
{2\pi\alpha^2 \left(2+\displaystyle\frac{1}{\tau}\right)}, 
\label{eq:Fp}
\ee
where $\alpha=e^2/(4\pi)$, $\beta=\sqrt{1-1/\tau}$, 
$~\tau={q^2}/(4M^2)$, $q^2$ is the squared invariant mass of the colliding 
pair, and $M$ is the proton mass. The effect of the Coulomb singularity 
of the cross section at the $\bar{p}p$ 
threshold is removed by the $\beta$ factor: 
$\beta$ $\rightarrow$ 0 for $q$ $\rightarrow$ $2M$, so that $\beta \sigma$ 
is finite and the effective form factor is expected to be finite at the 
threshold. 

\begin{figure}
\begin{center}
\includegraphics[width=8.5cm]{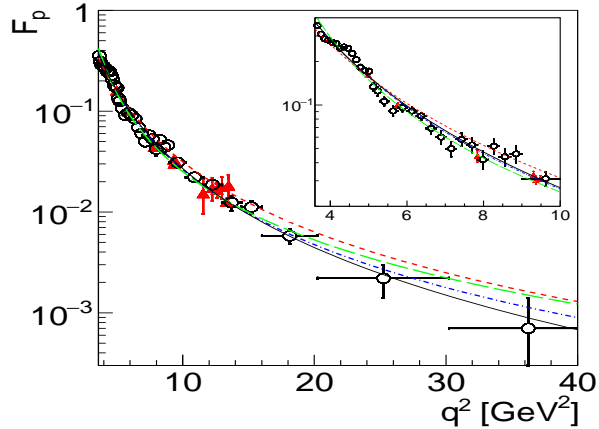}
\caption{Most recent data on the TL proton generalized FF as a function of $q^2$, from Refs. \protect\cite{Lees:2013xe,Lees:2013uta} (black open circles), 
Refs. \protect\cite{Ablikim:2005nn}  (red triangles),
together with the calculation from Eq. (\ref{eq:qcd}) (blue dash-dotted line), 
Eq. (\ref{eq:eak}) (red dashed line), Eq. (\ref{eq:BdT}) (green long-dashed line), and Eq. (\ref{eq:mpr}) (black solid line). 
}
\label{Fig:WorldData}
\end{center}
\end{figure}

This effective TLFF has been measured by several experiments for $q^2$ 
ranging from the threshold $(2 M_N)^2$ to about 36 GeV$^2$. The most recent and precise results  from the BABAR  \protect\cite{Lees:2013xe,Lees:2013uta} and BESIII collaborations  \protect\cite{PhysRevD.91.112004} are 
reported in Fig.~\ref{Fig:WorldData}.

These data have been fitted by some parameterizations. Here we report 
four of them, to give an idea of the general trend followed by the data 
and of the related ambiguities in extrapolations to the large $q$ 
region. 
Details about these fits and the best fit parameters 
can be found in our previous works \cite{Bianconi:2015owa,Bianconi:2016owb}. 
In the experimental papers before 
the year 2006, the function \cite{Ambrogiani:1999bh,Lepage:1979za}: 
\be
|F_{scaling}(q^2)|=\displaystyle\frac{\cal A}{(q^2)^2\log^2(q^2/\Lambda^2)},\ {\cal A}=40 ~\mbox{GeV}^{-4},\ \Lambda=0.45 ~\mbox{GeV}^{2}. 
\label{eq:qcd}
\ee  
was frequently used. The 
modification 
\be
|F_{scaling+corr}(q^2)|\ =\ 
\displaystyle\frac{\cal A}{(q^2)^2\left [ \log^2(q^2/\Lambda^2)+\pi^2 \right ]},\ {\cal A}=72~\mbox{GeV}^{-4},\ \Lambda=0.52 ~\mbox{GeV}^{2}. 
\label{eq:eak}
\ee
was suggested \cite{Shirkov:1997wi,Kuraev}
to avoid problems with ghost poles in $\alpha_s$. 
In Ref. \cite{Brodsky:2007hb} a pure rational form was 
proposed, with two poles of dynamical origin 
\be
|F_{T3}(q^2)|\ =\ 
\displaystyle\frac{\cal A}{(1-q^2/m_1^2) (2-q^2/m_2^2) },\ {\cal A}=1.56,\   m_1^2=1.5,\  \mbox{GeV}^{2}, \ m_2^2=0.77\ \mbox{GeV}^{2}.   
\label{eq:BdT}
\ee
The TLFF data from the BABAR collaboration \cite{Lees:2013xe,Lees:2013uta} 
extending from the threshold to $q^2$ $\approx$ 36 GeV$^2$, are steeper 
than the previous data, and are well reproduced by the following 
rational fit \cite{TomasiGustafsson:2001za}:
\be
|F_{BABAR}(q^2)|\ =\ 
\frac{\cal A}{(1+q^2/m_a^2)\left[1-q^2/0.71 \right]^2}, ~{\cal A}=7.7~\mbox{GeV}^{-4},~
\ m_a^2=14.8~\mbox{GeV}^2.
\label{eq:mpr}
\ee
where a $q^4$ asymptotic trend is not visible, although the data points at 
$q >$ 4 GeV present too large error bars to constrain the 
large-$q$ trend of a fit. 
For $q< $4 GeV the data also 
show oscillating 10 \% modulations around the previous fits. In our  
works \cite{Bianconi:2015owa,Bianconi:2016owb}, 
we have 
fitted the BABAR data with  
\be
F(p)\ \equiv\ F_0(p)\ +\ F_{osc}(p), 
\label{eq:diff}
\ee
where $p =p(q)$ is the relative three-momentum of the final hadron pair, 
$F_0(p)$ is any of the previous fits. Eqs. (\ref{eq:qcd}-\ref{eq:mpr}) are expressed in terms of $p(q)$, and 
the modulation term $F_{osc}(p)$ is parameterized as
\be
F_{osc}(p)\ \equiv\ A\ e^{- Bp}\ \cos(C p\ +\ D). 
\label{eq:eqdif}
\ee
The precise values of the parameters depend on which 
of the previous four fits is chosen as leading term $F_0$. 
A list of 
best fit values for all these cases is presented in 
\cite{Bianconi:2016owb}. 
In all cases $D\approx$0 and $A$ has magnitude $\simeq 0.1$. This means 
that the first oscillation is also a threshold enhancement, like 
those  found in $e^+e^-\rightarrow\bar{n}n$,  
$e^+e^-\rightarrow\bar{\Lambda}\Lambda$ and other 
production processes of neutral baryon 
pairs \cite{BESIII:2010ad,Pakhlova:2008vn,Ablikim:2006dw,Ablikim:2004dj,Bai:2003sw,Amsler:1994ah}.  

These near-threshold phenomena should disappear at 
large $q^2$, so that the data and their fits may converge to the simple 
quark counting rule: TLFF $\propto$ $1/q^4$, as predicted for the  
SLFF asymptotic \cite{Matveev:1973uz,Brodsky:1973kr}. 

This may be stated by using the same arguments of the SL 
case, that is by analyzing the dimensional structure of the 
matrix element \cite{Matveev:1973uz} or by assuming that at large $q$ the 
process is dominated by a PQCD hard core \cite{Brodsky:1973kr}, or 
by using analytic continuation at large $|q|$ from the SL to the TL 
sector (applying the Phragm\`en-Lindel\"of  theorem, see the discussion in 
\cite{TomasiGustafsson:2001za}). In all cases, 
the details of the soft part of the $\bar{p}p$ 
creation or annihilation process do not play a role. 
On the other hand, these features are expected to heavily affect the 
finite-$q$ deviations from the $1/q^4$ rule, and to determine the 
FF magnitude and phase. This has prompted several studies of the 
nonperturbative aspects of the TLFF. 
Some effects of bound-state gross features on PQCD calculations, 
leading to pre-asymptotic differences between TLFF and SLFF, 
were studied in Ref. \cite{Gousset:1994yh}, 
still within a largely perturbative scheme.  

Several detailed nonperturbative models for 
the nucleon or meson TLFF have been proposed:  
some derive from a unique analytic prediction valid both in the SL 
and in the TL region, other ones are more specific. There are approaches based on 
vector meson dominance  \cite{Bijker:2004yu, Adamuscin:2005aq} and 
dispersion relations  \cite{Belushkin:2006qa,Lomon:2012pn}. They  
give precise quantitative predictions for a large set of 
observables and have been applied \cite{Bianconi:2006wg,Bianconi:2006bb}, 
to simulate the feasibility of high-precision experiments 
including polarization observables and two-photon 
contributions \cite{Gakh:2005wa,Gakh:2005hh}. 
 
In \cite{deMelo:2003uk} 
a mixed approach to the pion TLFF 
is present, where VDM 
is applied at the level of photon-quark-antiquark vertex, but 
also a constituent quark loop and quark-pion couplings are 
present. In addition, a large number of poles is used, with 
parameters partly determined by phenomenology and partly by a 
dynamic model. Later on,  
non valence 4-constituent states have been added  \cite{deMelo:2005cy}. The approach based on 
AdS/QCD correspondence used in Ref. \cite{Brodsky:2007hb}, 
may be considered a pole-based model (see previous Eq. (\ref{eq:BdT})), 
although in this case the poles are not a starting assumption but rather 
the arrival point of a complex procedure. 

A distinguishing feature of the model presented 
in \cite{Kuraev:2011vq} is that it is built in spacetime, instead 
of momentum space. 
A large-$q$ suppression of the ratio of the electric to the magnetic 
FF in both the SL and TL sectors is suggested 
by a qualitative picture, where, in an intermediate stage of the 
hadron formation process, the reaction region is divided into 
a central region that is neutral from the color and flavor 
points of view, and a peripheral region where these properties 
are localized. At increasing $q$ this suppresses the overlap 
between the electric charge of the proton-antiproton pair,   
and the $1/q$-sized virtual photon. The suppression 
does not necessarily apply to the magnetic FF since a magnetic 
moment is not localized on the physical currents producing it. 

These models were targeted at the leading features of the 
data shown in Fig. \ref{Fig:WorldData}, the ``regular'' behavior 
reproduced by the above 
fits (\ref{eq:qcd}-\ref{eq:mpr}). The oscillations of Eqs. (\ref{eq:diff}-\ref{eq:eqdif}),  appearing as a periodic modulation, were interpreted in Refs. \cite{Bianconi:2015owa,Bianconi:2016owb} as an interference phenomenon in spacetime, with competition between 
processes involving well separated regions with 
different properties. In particular, regions 
closer to the $\gamma^*-q\bar{q}$ vertex would present 
regeneration properties for the $\bar{p}p$ wave function, 
while suppression of this state would occur in more 
peripheral regions. 
Starting from a different point of view, another fit to the  
oscillations of the TLFF was proposed by \cite{PhysRevD.92.034018} 
as a sum of independent structures like resonance poles and intermediate 
state thresholds. Interference in spacetime and poles in $q$ 
could be alternative ways to describe a similar mechanism: 
for the case of the pion TLFF, 
several oscillations regularly spaced in $q^2$ are predicted in 
the model by \cite{deMelo:2003uk}. Although they are due to the contribution of  
of many resonance states, these oscillations present a regularity 
pattern because of a unique dynamic model behind these resonances. 

The interpretation of the 
threshold enhancement is related to the oscillation problem, since 
the threshold enhancement can be seen as the 
first oscillation, although it seems especially evident 
in the TLFF of neutral baryons. 
The authors of Ref. \cite{Haidenbauer2014102} suggest that it is due to proton-antiproton 
strong interactions in low energy conditions. A different 
explanation was suggested by \cite{Baldini:2007qg}, in terms of local 
electric interactions between quarks and antiquarks of the two 
baryons. This is equivalent to a reciprocally induced  
electric polarization of the interacting spin-1/2 hadrons. Although 
nonstandard, the same mechanism has been used to explain the near-threshold 
rise of the inelastic antineutron cross sections in \cite{Bianconi:2014era}, and may find a justification in the  calculation of a neutron electric polarization induced by a  
strong external electric field due to QED  vacuum polarization 
terms \cite{Zimmer:2011yk}.  
\subsection{Aim of the present work}
Summarizing the previous discussion, the attempts to 
reproduce the non-perturbative aspects of TLFF data introduce 
complex and largely unexplored details of the hadron-pair formation 
process. 
Translating a model for TLFF into a spacetime picture of the hadron pair 
formation process is not immediate, however, since relativistic amplitudes 
are normally handled in momentum space, and the processes involving 
pair creation or annihilation do not have an intuitive  
nonrelativistic equivalent. The starting question of the 
present work is how one can translate data fits or models of TLFF into 
intuitive spacetime pictures of the forming or annihilating 
proton-antiproton system, similarly to what happened for SLFF. 

In the SL case, FFs
in the Breit frame ($q_0=$ 0, no energy transfer) may be 
interpreted in a standard nonrelativistic way, as Fourier 
space transforms of stationary charge and current distributions. 
The interpretation 
of the SLFF in terms of charge-current distribution has transformed   
a mathematical abstraction, 
that only experts of field theory may understand, into 
something that has a tangible meaning for a much broader audience. 


The SLFF interpretation in terms of a charge density cannot be extended to the TL case, 
since the photon time-like momentum 
can test time distributions of events, but not space 
distributions. In the CM frame of the $e^+e^-$ collision 
the photon has zero three-momentum (infinite space wavelength) 
so any effect related to space separation 
of electric charges is not detectable by it. Whatever is tested 
by the virtual photon, it must be a function $R(t)$ of the time 
deriving from an average over all the three-space. But, 
after a three-space average, the overall electric charge of the 
forming hadron-antihadron pair is equal to zero at any time. 
Of course, this concerns the "electric charge" in the classical 
electrodynamical sense, that is the source of an electromagnetic 
field. If we 
interpret the concept of  "charge'' as "photon-charge 
coupling'', we may think at $R(t)$ as an amplitude for 
creating charge-anticharge pairs at the time $t$. So, `"charge  
distribution'' can be understood as "distribution in time of 
$\gamma^* \rightarrow charge-anticharge$ vertexes''. 

In the following, we will examine in depth this idea, formalize the relation 
between $R(t)$ and the static space charge density $\rho(r)$ that is 
measured 
in the SLFF, and present some examples inspired by the phenomenology.

\begin{figure}
\begin{center}
\includegraphics[width=8.5cm]{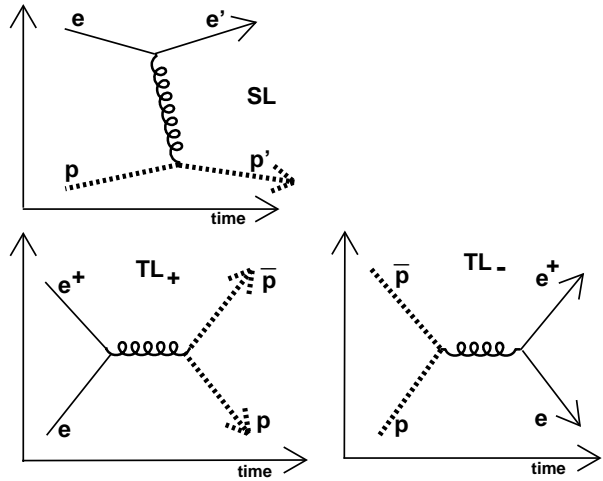}
\caption{
Feynman diagrams for reactions (\ref{eq:eq0}): labeled as SL in the figure, 
(\ref{eq:eq1}): TL$_+$, and (\ref{eq:eq2}): TL$_-$. In the one photon exchange approximation, electromagnetic 
FFs are functions characterizing the vertex coupling the virtual 
photon to the hadron current (thick dashed line in the figure). 
}
\label{Fig:reactions}
\end{center}
\end{figure}

\section{General definitions}
The relevant reactions for the extraction of SL and TL FFs are:  
\ba
&SL:\ &e^\pm + p \to e^\pm + p,
\label{eq:eq0}\\
&TL_+:\ &e^+ + e^- \to \bar p +p ,
\label{eq:eq1}\\
&TL_-:\ &\bar p + p \to e^+ + e^- .
\label{eq:eq2}
\ea
They are related by crossing symmetry and illustrated in  Fig. \ref{Fig:reactions}. 
Reaction (\ref{eq:eq0}) allows for measuring the FF 
in the spacelike (SL) kinematical region, corresponding to a virtual 
photon four-momentum $q_\mu$ with ${\vec q}^2 > q_0^2$. 
Reactions (\ref{eq:eq1}) and (\ref{eq:eq2}), allow for 
exploring the timelike (TL) FFs, more precisely, 
the processes (\ref{eq:eq1}) and  (\ref{eq:eq2}) are labeled $TL_+$ and $TL_-$ respectively.

\begin{figure}
\begin{center}
\includegraphics[width=8.5cm]{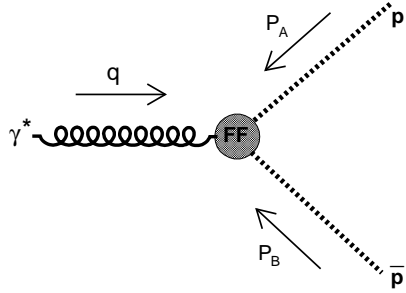}
\caption{
Subdiagram participating to the three reactions (\ref{eq:table1}-\ref{eq:table3}). Since they may be considered physical channels of the same reaction, all of them may be described 
by the same diagram and by the same amplitude  
by changing the values of the components of the 
four-momenta $P_A$, $P_B$, $q$, exploiting crossing symmetry. 
Formally, we consider these momenta 
as all entering. So, $q$ coincides with the physical four-momentum of 
the virtual 
photon in the channel $TL_+$ where a $\bar{p}p$ pair is created,  
while $q=-q'$, where $q'$ is the physical 
four-momentum of the virtual photon, in the reverse channel $TL_-$ 
where a virtual photon is produced 
by $\bar{p}p$ annihilation. Similar considerations apply to 
$P_A$ and $P_B$ (see Eqs. (\ref{eq:table1}-\ref{eq:table3}) for 
the correspondence between the formal arguments of the amplitude and 
the physical momenta). 
}
\label{Fig:diagram}
\end{center}
\end{figure}

We assume one-photon exchange, so in the following "form factor'' 
is meant as a factor renormalizing the hadron-virtual photon vertex, as in 
Fig. \ref{Fig:diagram}. Factorizing out the lepton part of the process and the virtual 
photon propagation, we will only consider the 
three-leg amplitude $A(q,P_A,P_B)$ 
describing the sub-processes of the reactions introduced above: 
\ba
SL&: &\ \gamma^*(q_\mu)\ +\ p(p_\mu) \rightarrow p(p_\mu') \label{eq:eqSL}\\
TL_+&:&\ \gamma^*(q_\mu)\ \rightarrow p(p_\mu')\ +\ \bar{p}(\bar{p_\mu}')\label{eq:TL+}\\
TL_- &:&\ p(p_\mu)\ +\ \bar{p}(\bar{p}_\mu)\ \rightarrow \gamma^*(q_\mu')  \label{eq:TL-}
\ea
The four-momenta $q^\mu$, $P_A^\mu$, $P_B^\mu$ appearing as $formal$ 
arguments of $A(q,P_A,P_B)$ 
are all incoming as in Fig. \ref{Fig:diagram},  
so that the different reactions are distinguished by the 
expression of $q$, $P_A$ and $P_B$ in terms of the $physical$ 
momenta $q$, $q'$, $p$, $p'$, $\bar{p}$, $\bar{p}'$ 
(that have a positive time component if they are timelike):  
\ba
&\gamma^* + p \rightarrow p'   \hspace{0.9truecm}  
(SL:\ |q_0|\ <\ |\vec q|)\  \hspace{1.8truecm}  
&P_A\ =\ p,\ P_B\ =\ -p', 
\label{eq:table1} 
\\ 
&\gamma^* \rightarrow \bar{p} + p\hspace{1truecm}  
(TL_+:\ |q_0|\ >\ |\vec q|,\ q_0\ >\ 0)\  
&P_A\ =\ -p',\ P_B\ =\ -\bar{p}'.
\label{eq:table2} 
\\ 
& \bar{p} + p \rightarrow \gamma^*\hspace{1truecm}    
(TL_-:\ |q_0|\ >\ |\vec q|,\ q_0\ <\ 0)\  
&P_A\ =\ p,\ P_B\ =\ \bar{p},\ q\ =\ -q'.
\label{eq:table3} 
\ea
whereas, in the TL region, two reciprocally inverse reactions are possible, 
corresponding to $\bar{p}p$ annihilation into (or creation from) 
a lepton-antilepton pair. 

It is important not to confuse the four-momenta $P_A,P_B$, as formal arguments  
of $A$, with their physical values $\pm p$, $\pm\bar{p}$ etc: 
the analytic continuation of $A(q,P_A,P_B)$ requires that 
this amplitude is described in terms of the same arguments in all the 
reaction channels and in the unphysical regions  
(so, $q_0<$ 0 in one of the two annihilation channels, and it is, in general, 
a complex variable). 
Being $A(q,P_A,P_B)$ invariant, it actually depends on $q$, $P_A$ and $P_B$ 
via their invariant products only, so these three four-vectors contain 
redundant information. However, in the following, we keep the formal 
dependence of $A$ on them. 

Here we distinguish between "resolvable'' and "unresolvable'' 
particles. A resolvable particle participates to a process with its internal 
structure, while an unresolvable particle is treated as a massive elementary 
particle. Both levels are present in the FF analysis. 
As an unresolvable particle, the photon-hadron current interaction 
takes place in a single vertex four-point $X_\mu$. The FF takes into account 
that at a resolvable level the 
photon-hadron interaction involves several variables $X_1^\mu,X_2^\mu,...$ 
associated to the internal hadron constituents. From now on we omit 
the tensor indexes and just write $X$, $X_1$ etc. 

Assuming a muon as a template for an unresolvable proton, 
the vertex matrix element for 
$\gamma(q)+\mu(p)\rightarrow \mu(p')$ is (using $\bar{u}\gamma^0= u^+$) 
\begin{align}
&A_{point\ SL}(q,p,p')\ =\ 
<\mu'| A_\nu(X) J^\nu(X)\ |\mu>\  =\ 
e \int d^4X\ 
e^{iqX}\ e^{-ip'X}\ e^{ipX}\ e_\nu\ \bar{u}(p') \gamma^\nu u(p)\ =\ 
\\
&=\ e \int d^4 X 
e^{iqX} e^{-ip'X} e^{ipX} 
\Big(
e_0\ u^+(p') u(p)\ -\ 
\vec{e}\ \bar{u}(p') \vec{\gamma} u(p)\Big) \ =\\ 
&=\   
\delta^4(q+p-p')\ \Big( T_{point\ charge}(q,p,p')\ -\ 
T_{point\ current}(q,p,p') \Big).  
\label{eq:a11_5}
\end{align} 
Exploiting that the amplitudes of the processes 
(\ref{eq:table2}) and (\ref{eq:table3})  
are analytic continuations of the amplitude of (\ref{eq:table1}), 
we can write Eq. (\ref{eq:a11_5}) in a form where it describes all 
these processes: 
\be
A_{point}(q,P_A,P_B)\ \equiv\ 
\delta^4(q+P_A+P_B)\ \Big( T_{point\ charge}(q,P_A,P_B)\ -\ 
T_{point\ current}(q,P_A,P_B) \Big),   
\label{eq:a11_5_gen}
\ee 
where, 
assigning to $q$, $P_A$, $P_B$ the values listed in 
Eqs. (\ref{eq:table1}-\ref{eq:table3}),  
we obtain the amplitudes for the corresponding reactions. 

FFs may be introduced as scalar functions that multiply the previous terms, 
or linear combinations of these terms:  
\ba
A(q,P_A,P_B)\ &\equiv& \ A_{charge}(q,P_A,P_B)\ -\ A_{current}(q,P_A,P_B)
\label{eq:a_split}
\\ 
&\equiv\ &\delta^4(q+P_A+P_B)\ 
\bigg(T_{point\ charge}(q,P_A,P_B)\ F(q)\ 
-\ T_{point\ current}(q,P_A,P_B)\ G(q) \bigg), 
\label{eq:Amp_start}
\ea
where now this amplitude describes processes involving proton and antiproton 
instead of muons. The scalar FFs 
$F(q)$ and $G(q)$ depend on $q_\mu$ via the scalar $q^2$ $\equiv$ 
$q_\mu q^\mu$ only. Alternatively, one may rewrite the hadron four-current in 
the Gordon form, insert $F_1$ and $F_2$ and next combine them into 
$G_E$ and $G_M$. However,  the adopted procedure 
is simpler, since it immediately highlights the term that is proportional 
to the charge density operator $u^+(p')u(p)$. We will not work on the 
other component in the following.  In the relevant reference frames (the Breit frame for the SL case and the CMS  for the TL case) 
$F(q)$ coincides with the electric form factor $G_E$. In an arbitrary frame, $F(q)$ is a linear combination of $G_E$ and $G_M$. 

Our further analysis \underline{only considers the form factor $F(q)$ 
associated with the charge term}. 
So our starting equation is:
\begin{align}
A_{charge}(q,P_A,P_B) 
\equiv\ \delta^4(q+P_A+P_B)\ 
T_{point\ charge}(q,P_A,P_B)\ F(q) 
\label{eq:Amp_charge_start}
\end{align}
\section{Fourier transform of the FF. 
SL-Breit and TL-CM cases, time density of photon-quark coupling.}
Following a suggestion from \cite{Kuraev:2011vq}, 
the key tool of our investigation is the four-dimensional Fourier 
transform
\begin{align}
F(q)\ =\ \int d^4 x e^{iqx} F(x). 
\label{eq:FT_start}
\end{align}


In the SL case, and in the Breit frame where $q^\mu=(0,\vec q)$, 
\begin{align}
&F_{SL,Breit}(q)\ =\ \int d^3 \vec x\ e^{- i \vec q \cdot \vec x}
\int dt F(t,\vec x) 
\ \equiv\ \int d^3 \vec x\ e^{- i \vec q \cdot \vec x}
\rho(|\vec x|)
\label{eq:FSLBR}
\\
&\rho(|\vec x|)\ =\ \int dt F(t,\vec x). 
\label{eq:rho}
\end{align}
where 
$\rho(|\vec x|)$ may be read as a static charge density. Here it appears 
as a time 
average over the Fourier transform $F(x)=F(t,\vec x)$. 

In the TL case, and in the CM frame ($\vec q $=0) 
\begin{align}
&F_{TL,CM}(q)\ =\ \int dt\ e^{i q t} \int d^3 \vec x F(t,\vec x) 
\ \equiv\ \int dt\ e^{i q t} R(t),
\label{eq:FTLCM}
\\
&R(t)\ =\ \int d^3 \vec x F(t,\vec x).
\label{eq:Rt}
\end{align}

It is evident that it is difficult, in absence of a model for 
the underlying $F(x)$ (that depends on both $\vec x$ and $t$), to 
find a simple relation between $\rho(\vec x)$ and $R(t)$, since they 
represent projections of the same distribution onto orthogonal 
subspaces. 

\section{General properties of F(x)}
Since we have required $F(q)$ to depend on the four-vector $q_\mu$ via $q^2$ 
only, $F(x)$ is constrained to have the 
form: 
\begin{align}
&{\vec x}^2\ >\ t^2:\ F_{out LC}(x_\mu)\ =\ f(x^\mu x_\mu).\\
&t^2\ >\ {\vec x}^2:\ F_{in LC}(x_\mu)\ =\ 
f_+(x^\mu x_\mu)\theta(t)\ +\ f_-(x^\mu x_\mu)\theta(-t)  
\label{eq:FF_form}
\end{align}
where we distinguish the ``in light-cone'' and the ``out of light cone'' 
components of $F(x)$. For $F_{out LC}(x_\mu)$, 
a $t\to -t$ asymmetry is forbidden by the 
requirement that symmetry properties of a scalar amplitude 
do not depend on the 
reference frame (a positive $t$ can be made negative by a proper  
Lorentz boost). Since a proper Lorentz boost cannot mix future  
and past light cones, the same constraint is not present on 
$F_{in LC}(x_\mu)$, that may be rewritten as 
\begin{align}
F_{in LC}(x_\mu)\ =\ 
1/2\ [f_+  + f_-]\ +\ 
1/2\ [f_+ - f_-] 
[\theta(t) - \theta(-t)]
\label{eq:FF_form2}
\end{align}
The last term is important since it leads to an imaginary part 
of $F(q)$ even if $F(x)$ is real. 

The $f_+-f_-$ term implies asymmetries between the reactions 
$\gamma^*$ $\rightarrow$ $\bar{p}$ $+$ $p$ and 
$\bar{p}$ $+$ $p$ $\rightarrow$ $\gamma^*$, supposedly associated 
with final/initial state interactions. These asymmetries 
are constrained by the T-reversal requirement that $|F(q)|$ is not changed 
by $q_0$ $\rightarrow$ $-q_0$ (proton-antiproton annihilation instead 
of creation), so the differences affect only phases. 


In absence of a physical model, there is no mathematical 
reason to prevent the TL form factor $F(q)$ from receiving contributions from  
the SL regions of $x_\mu$ and viceversa for the SL form factor. 
A simple example may confirm this: in the 
$(1+1)$-spacetime $(t,z)$ 
we may take $F(t,z)$ $=$ $\delta(z^2 - t^2 - 1)$, that is 
zero in the TL region $z^2$ $\leq$ $t^2$ including its borders. 
In the CM frame 
$R(t)$ $\equiv$ $\int dz F(t,z)$ $=$ $(1+t^2)^{-1/2}$. For real $t$, 
$R(t)$ admits a nonzero, 
real and regular (although analytically nontrivial) 
Fourier transform $F(q)$. 

On the contrary, within a physical model where relativistic causality 
is implemented,  the TL domains of $x$ are related to the TL domains 
of $q$. To demonstrate this, we need to discuss some of the physical content of 
$F(x)$. Up to now, $F(x)$ has just been introduced as 
the Fourier transform of a form factor. We now rewrite 
Eq. (\ref{eq:Amp_charge_start}), assuming a model where the virtual 
photon conversion into a proton-antiproton pair 
begins with the photon conversion into a quark-antiquark 
pair, and all the other steps of the process follow causally 
from this initial event. 



The amplitude describing how a free (anti)proton with momentum $p$  
splits into a Fock state of $N$ constituents is 
\begin{align} 
\psi(X_1,X_2,....X_N)\ \equiv\ e^{ipX}\Phi(x_1,x_2,...x_N)
\label{eq:separation}
\end{align}
where the four-factor $X_i$ is the spacetime position of the i-th constituent, 
$X$ is a linear combination of all the $X_i$, expressing the 
spacetime position of the proton as a whole (the unresolved proton) 
and the four-coordinates $x_i$ are internal four-coordinates relative to $X$:  
\ba
&&X\ =\ \sum w_i X_i,\ i=1,....N. 
\label{eq:coord1}
\\
&&x_i\ \equiv\  X_i-X.
\label{eq:coord2}
\\ 
&&\sum\ w_i x_i\ =\ 0,  
\label{eq:coord3}
\ea
where $w_i$ are weights that depend on dynamics (for example, on 
the longitudinal fractions or on the mass) within a given model. 

$\Phi$ is a fully relativistic amplitude, 
where each four-coordinate has an independent time dependence. 
$X$ is not the hadron CM  in 
nonrelativistic sense, since in Eq. (\ref{eq:coord1})
the positions of the partons are  
taken at different times. But, 
if the hadron current is not interacting with the environment,  
a four-coordinate $X$ must exist that makes the factorization of 
Eq. (\ref{eq:separation}) possible, because the $e^{i pX}$ term 
expresses the spacetime translation invariance of the 
(anti)proton as a whole, that is at unresolvable level. 

\begin{figure}
\begin{center}
\includegraphics[width=8.5cm]{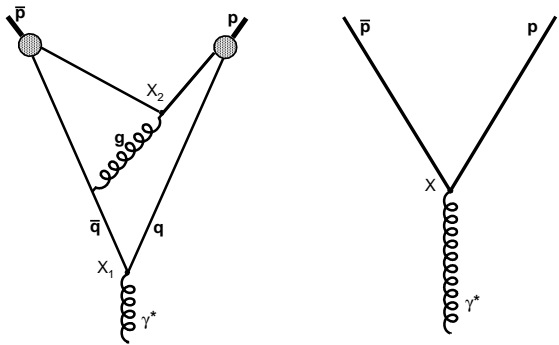}
\caption{
Left image: one of the possible chains of events that at resolved level 
lead to 
proton-antiproton formation from a virtual photon. In this figure the proton 
is schematized in a simplified form, as composed by a charged quark plus a 
neutral compact diquark. So $\gamma$ $\rightarrow$ $\bar{p}p$ requires 
at least 2 pair creation vertexes, in the four-points $X_1$ and $X_2$. 
Right image: the same process at unresolved level of analysis. Only  
one vertex is present in the four-point 
$X$, where the $\bar{p}p$ pair is directly created by the photon. 
The relation between $X$, $X_1$, and $X_2$ is determined by 
Eq. (\ref{eq:coord1}), that in this simple case will be of the form 
$X$ $=$ $w_1X_1$ $+$ $w_2X_2$.  
The corresponding geometry is represented in 
Fig. \ref{Fig:relative}. 
}
\label{Fig:particles}
\end{center}
\end{figure}

Let us first assume that, in the state of $N$ constituents, 
one quark only is charged. Its coordinate is $x_1$. 
Let $\psi'$ and $\Phi'$ 
refer to the final antiproton, and $\psi^+$ and 
$\Phi^+$ to the final proton. So we may rewrite 
Eq. (\ref{eq:Amp_charge_start}) for the process $\gamma^*$ 
$\rightarrow$ $\bar{p}p$ 
as:  
\begin{align}
&A_{TL,charge}\ =\ 
R_{point,charge}(q,p,\bar{p})\ e_1 \int dX_1 dX_2 .... e^{i iqX_1} 
\psi^+(X_1,X_2,...)\ \psi'(X_1,X_2,...)\ 
= \label{eq:invariance0}\\
&=\ R_{point,charge}(q,p,\bar{p})\ e_1 
\int dX e^{[i(q-p-\bar{p})X]} 
\int dx_1 e^{iqx_1} \int dx_2... \delta^4(\sum w_ix_i) 
\Phi^+(x_1,x_2,...)\ \Phi'(x_1,x_2,...)\ \equiv
\label{eq:invariance1}
\\
&\equiv\ 
R_{point,charge}(q,p,\bar{p})\ 
\delta^4(q-p-\bar{p})\  
\int d^4x_1 e^{iqx_1} F(x_1), 
\hspace{0.5truecm} x \equiv\ x_1. 
\label{eq:invariance2}
\end{align}
Here  $x_1$ is 
the four-point where the first quark-antiquark pair is created, 
while $x_2$ (or $x_3$, or other four-coordinates) 
could be the position where another quark-antiquark 
pair is created, not directly by the photon. 
A chain of processes leading from the pair created 
in $x_1$ to a second pair created in $x_2$ must exist. A 
standard PQCD example is a gluon radiated from the first quark that 
generates a second pair, as in Fig.\ref{Fig:particles}. 
The amplitude for processes like this may 
be absorbed inside $\Phi'(x_1,x_2,...)$ or $\Phi^+(x_1,x_2,...)$, 
or appear as a separate function describing the hard part of the 
process. 
Further functions may be introduced to consider later rescattering 
between the forming hadrons. 
This is not essential in the following, so the only functions we 
report explicitly are the hadron splitting functions. 

With more than one charged quark in a Fock state of $N$ constituents, 
$F(q)$ is at first order a sum over all the amplitudes where 
the photon directly interacts with one of these  
charges, so that in one amplitude 
$x$ $=$ $x_1$, in another one $x$ $=$ $x_2$ and so on. In addition, 
we must sum over Fock configurations involving different numbers 
of constituent partons or even intermediate state hadrons. 

These details concern the model one is applying, but in any case the 
structure suggested by 
Eqs. (\ref{eq:coord1},\ref{eq:invariance2}) will be present. 
We will find a four-coordinate $X$ 
representing the point where the photon creates the $unresolved$ 
proton-antiproton pair. This coordinate leads to the 
momentum-conserving $\delta^4$ function, and has no other role. 
Indeed, being $e^{iq_\mu X^\mu}$ the wave function of the photon, all the 
spacetime points are perfectly equivalent for this creation. 
The coordinate separation and the introduction of relative coordinates 
in Eq. (\ref{eq:coord2}) implies that 
the form factor is calculated by implicitly assuming that 
the unresolved proton-antiproton pair is created in the origin. 

At resolved level, in the diagram where the i-th quark-antiquark 
pair is the active pair directly created by the photon,  the argument 
$x$ of the form factor 
is the four-position $x_i$ of this pair creation with respect to the 
origin. 

Let us again consider for simplicity the case where only the 
quark-antiquark pair ``1'' is charged. $R(t)$ is an integral of 
the form $\int d^3\vec{x} \int d^4 x_2 ... $. In a model 
for $e^+e^-$ $\rightarrow$ $\bar{p}p$ 
where all the events $x_2$, $x_3$, ... are causally consequent to 
the first pair creation in $x$ $=$ $x_1$, all the four-points $x_2$, 
$x_3$ .... must be in the future light cone of $x$, and 
$t$ $=$ $t_1$ is the most negative of all the involved times 
$t_1$, $t_2$... $t_N$. Because 
the origin is an average of all the $x_i$ with positive coefficients 
$w_i$, 
the origin is in the future light cone of $x$ $\equiv$ $x_1$. So 
$t$ is negative, and $x$ $=$ $(t,\vec x)$ is in the past light cone 
of the origin. In the reverse process 
$\bar{p}p$ $\rightarrow$ $\gamma^*$, 
the same logic implies $t$ $>$ 0, and $x$ is in the future light cone 
of the origin. 
\begin{figure}
\begin{center}
\includegraphics[width=8.5cm]{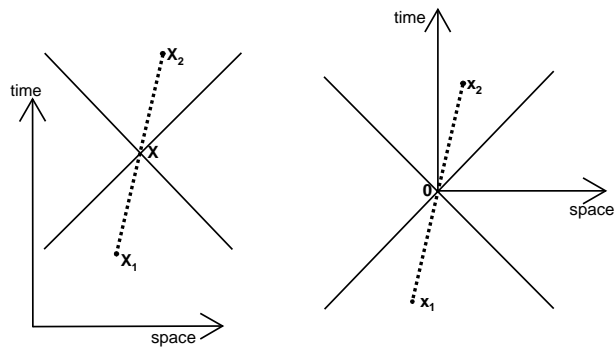}
\caption{
Absolute (left image) and relative (right image) coordinates 
for the chain of 
events leading to $\bar{p}p$ formation from a virtual photon 
as shown in Fig. \ref{Fig:particles}. Left image: 
according to 
Eqs. (\ref{eq:coord1}), $X_1$, $X$ and $X_2$ lie along a straight line,  
that is represented as a thick dashed line in the figure. This 
line does not correspond to any physical particle, we just use  
it to highlight the relative position of the three points. The 
continuous thin straight lines at 45$^o$ and 135$^o$ 
represent the light cone of $X$. Right image: 
the same geometry of the left image, but using the relative 
four-coordinates $x_1$ and $x_2$ introduced in Eqs. \ref{eq:coord2}. With 
this transformation, $X$ becomes the origin. The four-coordinate 
$x$ that is argument of the space-time form factor 
$F(x)$ coincides with $x_1$, the 
four-point where the photon creates the first 
quark-antiquark pair. 
}
\label{Fig:relative}
\end{center}
\end{figure}

The previous Eqs. (\ref{eq:invariance0}-\ref{eq:invariance2})  
could be repeated for the 
SLFF. In this case however, $x$ would not lie in the (past or future)  
light cone of the origin. This means that although $\Phi^+(x_1,x_2,...)$ 
may represent a final proton with the same four-momentum in both the SL 
and the TL cases, the identity between 
$\Phi_{SL}^+(x_1,x_2,...)$ and $\Phi_{TL}^+(x_1,x_2,...)$ must be meant 
in analytic continuation sense. Measures in the SL sector produce a 
knowledge on $\Phi(...)$ that requires an extrapolation, to be applied 
to the TL sector. The same must apply to $F(x)$. 
\section{Examples}
The simplest examples approximate the proton as 
``single charged active quark plus neutral spectator diquark''. As above indicated, let the origin $X=(0,0,0,0)$ 
be the four-point where the unresolved $\bar{p}p$ pair is created. 
Let $(t_1,\vec x_1)$ be the point where the initial active 
quark-antiquark pair is created, and $(t_2,\vec x_2)$ the point  
where the spectator-antispectator pair is created. Then, 
following Eq. (\ref{eq:coord2}),  we have 
\begin{align}
x\ \equiv\ x_1\ =\ (x_1-x_2) w, \hspace{0.5truecm} w\ >\ 0.  
\label{eq:coordb1}
\end{align}
For example, in the symmetric case we have $x_1=-x_2$ and 
$x=(x_1-x_2)/2$. In general, $w$ may depend on parton 
masses and dynamics. Here the only relevant things are the following: 

Causality implies that $t_1<t_2$, and since the weight coefficient 
$w$ is positive the origin is somewhere on the straight line joining 
$x_1$ and $x_2$. Since $x_2$ is in the future light cone of $x_1$, 
the origin is in the future light cone of $x=x_1$, and $t< 0$. 

In the initial examples we violate T-symmetry assuming that $F(x)$ 
is nonzero only for negative times (that describes proton-antiproton 
creation but not annihilation). Next we add the reverse process  
piece. 
\subsection{Case 1. Homogeneous distribution for positive times} 
We assume that after the initial quark-antiquark creation, 
the creation of the complete proton-antiproton system is possible at 
any time with equal probability if this happens inside the future 
light-cone of the first event. 
We don't 
know how this probability is spatially distributed, but the integral over 
all the space is time-independent and we 
fix it to 1 at any given time. 
Since the unresolved $\bar{p}p$ pair is created for $t=0$, the 
condition "$\bar{q}q$ pair created before $\bar{p}p$ pair"  just means  
$t<0$.  
\begin{align}
R(t)\ =\ \theta(-t), 
\label{eq:theta}
\end{align}

\begin{align}
F(q)\ =\ \int e^{iqt} \theta(-t)\ =\ {\pi \over {\epsilon - iq}}, 
\label{eq:FF1}
\end{align}
with infinitesimal $\epsilon$. 
\subsection{Case 2. Exponential damping} 
Common sense suggests that either the spectator pair and the complete 
proton-antiproton system are created soon after the active pair, 
or the process will lead to independent fragmentation of the 
initial quark and antiquark. So 
it is more realistic to generalize Eq. (\ref{eq:theta} )
to 
\begin{align}
R(t)\ =\ \theta(-t) e^{-a|t|}. 
\label{eq:thetaexp}
\end{align}
that suppresses the probability of the creation of an exclusive hadron pair for 
$|t|\gtrsim 1/a$. This leads 
to 
\begin{align}
F(q)\ =\ {\pi \over {a - iq}}\ =\ 
{{a \pi} \over {a^2+q^2}} + i{{q \pi} \over {a^2+q^2}} , 
\label{eq:FF2}
\end{align}
where 
the difference with respect to the previous case is that $a$ is finite. 
\subsection{Case 3. Monopole-like shape}
As observed in a previous section, $F(x)$ must 
be nonzero both in the future and in the past light-cone, 
to describe  
both $\bar{p}p$ creation or annihilation. These terms should be 
time-symmetric, apart for a possible phase difference. 
We sum two terms like the 
previous one, corresponding to positive and negative $t$. Taking  
them with the same phase, we get a monopole-like distribution, with 
the correct 
asymptotic of the form factor of a two-constituent hadron: 
\begin{align}
R(t)\ =\ \theta(t) e^{-at}\ +\ \theta(-t) e^{at}\ =\ e^{-a|t|}
\label{eq:dipole1}
\end{align}
\begin{align}
F(q)/\pi\ =\ {1 \over {a - iq}}\ +\ {1 \over {a + iq}}\ =\ 
{{2a} \over {a^2 + q^2}}. 
\label{eq:FF3a}
\end{align}
The $1/a$ parameter has the meaning 
of a formation time. In this simple two-constituent model of the proton, 
we have two meaningful pair creation vertexes at times $T_1$ and $T_2$. 
This implies one relative time $t$, that according to 
Eqs. (\ref{eq:coord1},\ref{eq:coord3}) has the magnitude of $t_1-t_2$ 
(for example, in a symmetric model $t=(t_1-t_2)/2$). 
For $|t| \gg 1/a$, $R(t)$ is very small. This means that either 
the second pair is formed within $1/a$, or the initial pair will 
produce two separate hadron showers.  

When $q \approx$ some quarkonium mass, the scale of this time may 
expand to the time life of a resonance: the initial pair may form a 
long-lived state, and the second pair has more time to be formed. 
This is discussed in detail below. 
As it is, Eq. (\ref{eq:FF3a}) corresponds to a zero-mass resonance of 
width $a$. 

The above monopole form with its $R(t)$ counterpart   
contains two properties of general character: (a) a correct $1/q^2$  
asymptotic for the formation of a hadron pair when each hadron is 
formed by 2 constituents, (b) the presence of a time cutoff $1/a$, 
meaning that the formation of the full hadron pair and of the first 
quark-antiquark pair cannot be too far in time. 
\subsection{Case 4. Resonance-like, space and time parameters}
Eq. (\ref{eq:FF3a}) may be written as 
\begin{align}
F(q)/\pi\ =\ i \Big({1 \over {q + ia}}\ -\ {1 \over {q - ia}} \Big). 
\label{eq:FF3b}
\end{align}
The $simplest$ way to have poles with nonzero mass is to 
substitute $q \to q-M$ 
leading to a Lorentzian (not Breit-Wigner) resonance shape: 
\begin{align}
F(q)/\pi\ =\ i \Big({1 \over {q - M + ia}}\ -\ {1 \over {q - M - ia}} \Big)\ 
\propto\ 
{1 \over {(q-M)^2} + a^2}.  
\label{eq:FF4a}
\end{align}
This shape describes, for example, 
the stationary response of a classical damped oscillator 
to an external periodic force. 
By Fourier transform we get  
\begin{align}
R(t)\ \propto\ \ e^{iMt}\ e^{-a|t|}
\label{eq:dipole2b}
\end{align}
Since 
a Fourier transform is a sum with homogeneous weight over all the 
frequencies, the previous $R(t)$ is the response 
of a classic damped oscillator to an instantaneous external force 
of the form $\delta(t)$ (Eqs. (\ref{eq:FF4a}, \ref{eq:dipole2b}) are 
the frequency and time Green functions of that problem).  

Although a classical oscillator presents several similarities 
with some quantum systems, it has not the 
problem of the negative-energy solutions of the relativistic 
wave equation. 
We should remind that here $q$ means $q_0$ (the time component 
of the 4-vector $q_\mu$) and not $\sqrt{q^2}$, so it  
may be negative 
(we are in the CM frame where $\vec q = 0$). 
Because of the relativistic particle-antiparticle 
symmetry, to each pole with $q_0=M+ia$ corresponds  a pole with 
$q_0 =-(M+ia)$ that  describes the corresponding negative-energy states. 
With only positive $Re(q_0)$ poles, we are back to the situation of the 
first two examples of this section, where 
$F(q)$ describes the $\bar{p}p$  
creation process, but not the annihilation one. Indeed, by closing 
the integration path on the upper or lower half of the complex plane 
the Fourier transform returns us an $R(t)$ containing $\theta(\pm t)$. 
The two poles must be exactly opposite, so that the situation 
is unchanged if the physical photon energy $q_0'=-q_0$ of the 
$\bar{p}p$ annihilation channel is used instead 
of $q_0$ to describe the amplitude. 

A Breit-Wigner $probability$ distribution contains all the four poles 
$q_0$ $=$ $\pm (M \pm ia)$. The corresponding amplitude is 
\begin{align}
F_\pm(q)\ \propto\ 
{1 \over {(q^2-M^2)} \pm iMa}  
\label{eq:FF4b}
\end{align}  
where we may imagine several combinations of $F_\pm(q)$ composing 
a form factor. For example 
\begin{align}
F(q)\ \propto\ 
F_+(q) + F_-(q)
\label{eq:FF4c}
\end{align}
corresponds to 
\begin{align}
R(t)\ \propto\ \ cos(Mt)\ e^{-a|t|}.
\label{eq:dipole2c}
\end{align}
and  
gives $F(q)=O(1/q^2)$ at large $q$, as expected for the 
two-constituent hadron we are working with. 

\begin{figure}
\begin{center}
\includegraphics[width=8.5cm]{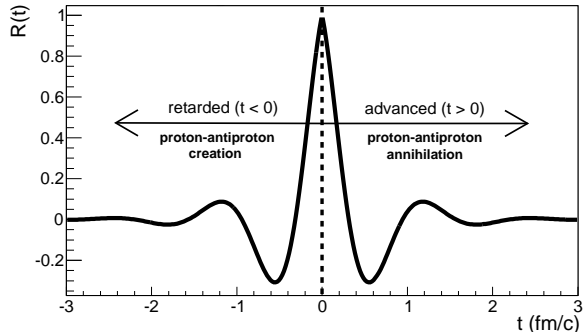}
\caption{
$R(t)=\cos(Mt)e^{-a|t|}$, with $M= 1$ GeV, $a=0.4$ GeV, 
as in Eq. (\ref{eq:dipole2c}). The retarded ($\bar{p}p$ creation)
and advanced (annihilation) contributions 
of Eq. (\ref{eq:dipole2d}) are distinguished. 
}
\label{Fig:cos1}
\end{center}
\end{figure}

$F_+(q)$ and $F_-(q)$ contain respectively one pole from the  
the $\bar{p}p$ creation and one pole from the $\bar{p}p$ 
annihilation process. 
With arguments similar to those following Eq. (\ref{eq:dipole2b}), 
we may say that Eq. (\ref{eq:dipole2c}) sums two contributions, that 
may be highlighted by writing 
(see Fig.\ref{Fig:cos1}) 
\begin{align}
R(t)\ \equiv\ R_{creation}(t)\theta(-t)\ +\ R_{ann}(t)\theta(t).
\label{eq:dipole2d}
\end{align}
One of the two pieces 
describes the process in the $\bar{p}p$ creation channel, and it has 
the same form as the retarded response of a classical bound and damped 
oscillating system to a $\delta(t)$-shaped external perturbation. 
The other one has the same meaning, in the $\bar{p}p$ annihilation 
channel. Analytically, it may be also read as an unphysical 
advanced response in the creation process. 

We know that the tail of a resonance  
may be much more complicated than this, and pole-based models of FFs 
\cite{Bijker:2004yu,Belushkin:2006qa,Iachello:2004aq} are 
more sophisticated than the above Lorentz and BW 
examples. However, the BW example 
contains the basics to remark a few points. First, $two$ dimensional 
and scaling-violating parameters  
appear, corresponding to the pole mass and width. For obvious reasons, 
in the SL analytic continuation $q^2\to -q^2$ 
the leading parameter expressing how  
a charge distribution decreases with the distance is the pole mass. 
In the TL case this mass is associated with the frequency of 
the oscillation in time of the 
underlying photon-quark-antiquark coupling. 
The parameter that 
tells us how fast is the decrease in time of the probability of 
the formation of the hadron pair is the pole width. Taking into account 
that fast-decaying hadron resonances have mass $\sim$ 1 GeV, and 
standard width in the range 0.1-1 GeV, we expect for $R(t)$ a shape 
like in Fig. \ref{Fig:cos1}, with a small number of visible 
oscillations. If the pole had zero width 
the oscillation would continue forever, like in the first example 
of this section where $a$ was infinitesimal leading to $R(t)$ $=$ 
$\theta(t)$. This would not prevent from having a finite charge 
radius in the SL measurement given by $<r> \sim 1/M$. 
The SLFF would appear as a monopole $1/(|q^2|+M^2)$. 
\subsection{Case 5. Several spectators: 
dipole and asymptotic $1/q^{2(n-1)}$ behavior}
A nucleon is made of three constituents in its basic valence state, 
possibly more in temporary fluctuations. Because of the 
valence structure, 
for the nucleon FF we expect a $1/q^4$ law at large $q$, and more 
in general a $1/q^{2(n-1)}$ law if the produced hadrons are made 
of $n$ compact constituents. Since this behavior does not 
depend on the relative wave function or interaction of these 
constituents, we would like to identify a mechanism that leads to 
the correct asymptotic form, whichever these details may be. 

We may use the Fourier transform property of 
convolutions: 
\begin{align}
F_1(q)F_2(q)\ =\ 
F.T.\Big[R_1(t)*R_2(t)\Big], 
\label{eq:conv1}
\end{align}
where 
\begin{align}
\Big[R_1(t)*R_2(t)\Big]\ \equiv\ \int d\tau R_1(\tau)R_2(t-\tau).
\label{eq:conv2}
\end{align}
So a function like 
\begin{align}
F(q)\ \propto\ {1 \over {(a^2 + q^2)(b^2 +q^2)}}, 
\label{eq:conv3}
\end{align}
that 
presents the required asymptotic trend, is the Fourier transform 
of 
\begin{align}
R(t)\ =\ \int d\tau\ 
e^{-a|t-\tau|}\ e^{-b|\tau|}. 
\label{eq:conv4}
\end{align}
This 
contains the required statistical properties. In a three-constituent  
Fock state the proton has two internal (four-dimensional) degrees 
of freedom. One of the two convoluted terms has the same role and meaning 
it had in the previous two-constituent case, and is associated to the 
degree of freedom that is directly probed by the virtual photon. 
The other term represents a decaying correlation between the 
active and a spectator degree of freedom. 
Being dominated by simple valence configurations, the 
large-$q$ behavior will derive from  a sum of 
three terms like Eq. (\ref{eq:conv4}). In each term one of the three valence quarks 
plays the role of active quark. 

\begin{figure}
\begin{center}
\includegraphics[width=8.5cm]{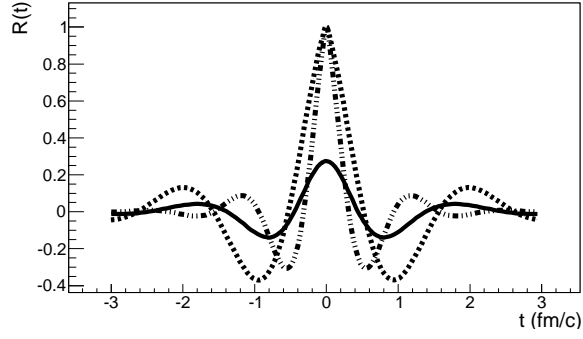}
\caption{
Dotted line: $R_1(t)=\cos(M_1t)e^{-a|t|}$, with $M= 1$ GeV, $a=
0.4$ GeV; thick-soft double-dotted line: 
$R_2(t)=\cos(M_2t)e^{-b|t|}$, with $M_2 = 0.6$ GeV, $b=0.1$ GeV;  
continuous line: convolution $[R_1(t)*R_2(t)] = \int d\tau R_1(\tau)R_2(t-\tau)$ according to Eq. (\ref{eq:conv2}). 
\label{Fig:cos_double}
}
\end{center}
\end{figure}

In Fig. \ref{Fig:cos_double} we show 
an example of convolution with $R_1(t)$ and $R_2(t)$ of resonance 
type (see Eq. (\ref{eq:dipole2c})). 
The final shape depends (even at qualitative level)  
on the parameters of the convoluting $R_1$ and $R_2$, but 
some rules are simple: If the decay times of $R_1$ and $R_2$ 
are different, $R_1*R_2$ coincides at large $|t|$ with the 
one between $R_1$ and $R_2$ with the longer lifetime. 
$R_1*R_2$ may decay for two 
reasons: (a) because the oscillations of $\cos(M_1t)$ and $\cos(M_2t)$ 
acquire opposite phase (for $t\approx\pi/(M_1-M_2)$), (b) 
because $t >1/a_{long}$, where ``long'' refers to the 
longer-life pole. 
So the decay time of the convolution is determined by the largest 
between $|M_1-M_2|$ 
and the width of the longer-life pole. If the process is dominated by 
standard hadron poles like $\rho$, $\omega$ the decay time is 
of magnitude 1/(200 MeV) $\sim$ 1 fm. Narrow large-mass poles 
could lead to much more unpredictable effects. Since the poles 
entering the 
convolution are poles of quark-antiquark states, they can also be 
poles of the full proton-antiproton system. 

\begin{figure}
\begin{center}
\includegraphics[width=8.5cm]{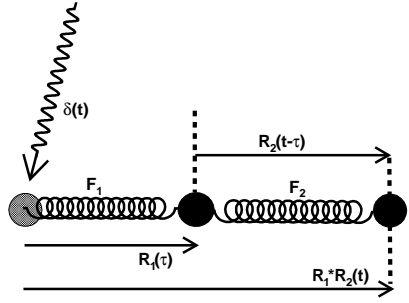}
\caption{
Sequence of oscillators corresponding to the response function 
of Fig. \ref{Fig:cos_double}. 
In equilibrium both the black masses overlap with the grey circle. They can 
move horizontally under the action of 
elastic forces $F_1$ and $F_2$, graphically represented as springs, 
or of external forces. 
When any of these masses is subject to an instantaneous external impulse 
$\propto$ $\delta(t-t_0)$ at the time $t_0$, its later displacement from 
the equilibrium position is described by the Green response function 
$R_i(t-t_0)$. 
A short impulse by an external force $\delta(t)$ at $t= 0$ causes 
the displacement $R_1(\tau)$ of the first mass at the later time $\tau$. 
The displacement of the first mass acts as an external force on the 
second mass, and may be decomposed into short impulses: 
$R_1(\tau)=\int d\tau' R_1(\tau') \delta(\tau-\tau')$. 
Since each short impulse 
at the time $\tau$ produces a response $R_2(t-\tau)$ of the second mass, 
the resulting displacement of the second mass is 
$\int d\tau R_2(t-\tau)R_1(\tau)$. 
\label{Fig:molle22}
}
\end{center}
\end{figure}

The dynamical meaning of the convolution in Fig. \ref{Fig:cos_double}
is described in Fig. \ref{Fig:molle22}. 
As observed after Eqs. (\ref{eq:dipole2b}) and 
(\ref{eq:dipole2d}), $R(t)$, when derived from a Lorentz or 
Breit-Wigner form,
corresponds to the response of a classical 
damped oscillator to a $\delta(t)$ external perturbation. The 
convolution structures of Eq. (\ref{eq:conv2})
describe the response of a chain of two oscillators, 
where one end of the chain is directly under the strain of the 
virtual photon. 

In sub-asymptotic conditions more 
degrees of freedom could play a role. These terms would 
imply a longer chain of convolutions. 
For example, with four constituents we would  
have 
\begin{align}
R(t)\ =\ 
\Big[[R_1(t)*R_2(t)]*R_3(t)\Big] 
\label{eq:conv5}
\end{align}
leading to a form factor that empirically could 
appear as a product of monopoles 
\begin{align}
F(q)\ \propto\ 
{1 \over {(q^2 \pm a^2)(q^2 \pm b^2)(q^2 \pm c^2)}}.  
\label{eq:conv6}
\end{align}
where the sign (in the TL channel) is negative if the mass is larger  
than the width of a pole, positive in the opposite case. 
A three-pole structure would be found in a process like 
$e^+e^-\rightarrow\bar{p}n\pi^+\rightarrow\bar{p}p$ 
where three quark-antiquark creation 
vertexes $x_1$, $x_2$, $x_3$ are needed to create the intermediate 
state. 
For example, the data from the BABAR collaboration \cite{Lees:2013xe,Lees:2013uta} are well fitted by 
Eq. (\ref{eq:mpr}) that has the sub-asymptotic form 
$F(q) \propto 1/(q^2+a^2)(q^2-b^2)^2$. 
\subsection{Case 6. Oscillating modulations, and delayed or advanced terms}
If we have the sum of two contributions of equal shape 
\ba
R(t)\ &=&\ R_0(t)\ + a R_0(t-b),\ a \ll 1,\\ 
F(q)\ &=&\ F_0(q) [1 + a e^{ibq}],
\label{eq:oscillation1}
\ea
because 
of a known property of the Fourier transforms: $F.T.[G(t-b)]= e^{iqb}F.T.[G(t)]$. 

We expect a similar phenomenon if the second distribution is not exactly 
identical to the first one, but is similar.   
For example, 
$R_0(t)$ could have a peak in $T$, and $R_1(t)$ a similar peak in $T-b$. 
This would  lead to a periodic modulation. 

The oscillating modulation discussed 
in \cite{Bianconi:2015owa,Bianconi:2016owb} 
however, shows a periodic pattern with respect to the final state 
hadron relative momentum, rather than to $q$. So, that phenomenon 
requires a more complex explanation, where the role of the final 
state kinematics is more explicit. 
\section{Conclusions}
We have explored a scheme where the TL hadron FF 
is interpreted as an amplitude for the 
distribution in time of the quark-antiquark pair creation vertex. 
This is the timelike counterpart of the known interpretation of the spacelike 
form factor as the Fourier transform of a classical charge distribution. 

Exploiting analytic continuity between the physical reactions where 
both FFs are measured, these are considered to be 
the analytic continuation of a unique function $F(q)$. For 
real values of the components of $q_\mu$, $F(q)$ is assumed to be the 
four-dimensional Fourier transform of a unique function $F(x)$, that is 
$F(q)\equiv \int e^{iq_\mu x^\mu} F(x)$. 

Giving to $q_\mu$ the spacelike and timelike components $(0,\vec q)$ 
and $(q,0)$, we get $F_{SL}(q) = \int d^3 \vec x \rho(\vec x)$, 
and $F_{TL}(q) = \int d t R(t)$, where $\rho(\vec x)=\int dt F(x)$, and $R(t)=\int d^3 \vec x F(x)$. 
So the distributions  
that are tested by the virtual photon wave are projections onto 
orthogonal one-dimensional and three-dimensional spaces of the same underlying 
function $F(x)$. 

We have next explored the main properties of the function $F(x)$. 
The contributions to the 
timelike form factor appearing in the reactions of proton-antiproton 
creation and annihilation originate from those $x$ that lie in 
the future and past light cones of the origin. The former contributes 
to the $e^+e^- \rightarrow \bar{p}p$ reaction, the latter 
to the reverse process. A phase asymmetry between the values of 
$F(x)$ in the two light cones is allowed by general invariance rules. 
This in principle permits an imaginary part to be present in $F(q)$ 
even if $F(x)$ is real. 

Next we have presented some simple examples for possible $R(t)$ functions 
with consequent form factors. These were not models, but rather the simplest possible functions 
presenting realistic phenomenological features: a dimensional parameter 
associated with the hadron pair formation time, the expected large $q$ power 
counting behavior, and interference phenomena. 

In conclusion, the present interpretation of FFs in the time-like region
highlights the spacetime meaning of these fundamental quantities, and 
relates the static charge density features with the time evolution  
properties of the hadron pair formation. 

This interpretation will help understanding high precision data 
expected to come from future measurements. Experimental programs at all existing and planned hadron facilities are on going or foreseen, for example at Mainz (Germany), JLab (USA) in the SL region, and, in the TL region, at VEPPII (Russia), BESIII at BEPC2 (China) and at the future antiproton facility PANDA at FAIR (Germany).




\end{document}